# Termination of Single Crystal Bi$_2$Se$_3$ Surfaces Prepared by Various Methods

Weimin Zhou, Haoshan Zhu, Jory A. Yarmoff*
*Department of Physics and Astronomy, University of California, Riverside CA 92521*

**Abstract**

Bismuth Selenide (Bi$_2$Se$_3$) is a topological insulator with a two-dimensional layered structure that enables clean and well-ordered surfaces to be prepared by cleaving. Although some studies have demonstrated that the cleaved surface is terminated with Se, as expected from the bulk crystal structure, other reports have indicated either a Bi- or mixed-termination. Low energy ion scattering (LEIS), low energy electron diffraction (LEED) and x-ray photoelectron spectroscopy (XPS) are used here to compare surfaces prepared by *ex situ* cleaving, *in situ* cleaving, and ion bombardment and annealing (IBA) in ultra-high vacuum (UHV). Surfaces prepared by *in situ* cleaving and IBA are well ordered and Se-terminated. *Ex situ* cleaved samples could be either Se-terminated or Bi-rich, are less well ordered and have adsorbed contaminants. This suggests that a chemical reaction involving atmospheric contaminants, which may preferentially adsorb at surface defects, could contribute to the non-reproducibility of the termination.

___________
*Corresponding author, E-mail: yarmoff@ucr.edu

## I. Introduction

Topological insulator (TI) materials have attracted much attention due to the topological surface states (TSS) that make the surface conductive in two-dimensions while the bulk remains insulating [1-3]. The spins of the surface carriers are locked to their momentum making these materials robust against non-magnetic defects because the spins have to be flipped for the carriers to change direction when they backscatter [4-6]. The unique properties of TIs have made them promising materials for next generation devices based on novel approaches such as spintronics and quantum computation [2,4,6,7].

$Bi_2Se_3$ is the prototypical TI material, but there are conflicting reports about the actual surface composition. $Bi_2Se_3$ is a layered material with a basic quintuple layer (QL) building block consisting of five atomic layers ordered as Se-Bi-Se-Bi-Se [8]. The QL's are attached to each other by a weak van der Waals force so it would be expected for the surface to naturally cleave between QL's. An intact QL would be terminated with Se atoms and have Bi in the second layer. Such a Se-termination has been widely demonstrated for samples cleaved *in situ* under vacuum [9-11]. A previous study from our group had found a Bi-termination, however, and included density functional theory (DFT) calculations which showed that termination by a complete Bi bilayer is energetically more favorable than the Se-termination [12]. For surfaces cleaved *ex situ* in air, the results in the literature are more contradictory. Kong, et al. found that the surface became oxidized after exposure to air for less than 10 s [13], while Atuchin, et al. found an inert Se-terminated surface even after a month-long exposure to air [14]. Edmonds, et al. used x-ray photoelectron spectroscopy (XPS) to show that metallic Bi forms at the surface after a 5 min exposure to air, but the surface would either oxidize or revert back to a Se-termination after extended exposure [15]. Hewitt, et al. found that both Bi-rich and Se-terminated surfaces are possible after *ex situ* cleaving with carbon tape, and that the probability for obtaining a Bi-rich surface is 48%, but drops to 25% if the sample is stored in vacuum [9]. Note that a sample with metallic Bi at the surface is labeled here as Bi-rich, instead of Bi-terminated, as it's not always clear if the excess Bi forms an epitaxial overlayer, amorphous film or isolated metallic islands. Coelho, et al. demonstrated the coexistence of a Bi bilayer termination and a Te-termination on a $Bi_2Te_3$ surface prepared by ion bombardment and annealing (IBA) [16]. $Bi_2Te_3$ has the same crystal structure as $Bi_2Se_3$ and is also a TI.

A critical issue for $Bi_2Se_3$ is that the Fermi level often resides inside the conduction band making the bulk metallic, which degrades the material's usefulness as a TI because it becomes difficult to isolate the TSS from the bulk states [3,17-20]. It has also been found that this Fermi level movement can occur slowly over time after cleaving, which is the so-called "aging effect" [19]. This has generally been attributed to a natural n-type doping by defects in the surface region, such as Se vacancies, but the specific nature of these defects has not been clearly demonstrated. The Fermi level movement and natural doping are likely related since electronic properties depend on atomic structure and composition. Additional investigations of the surface structure are needed to understand the underlying cause of the Fermi level movement and provide solutions that preserve the electronic properties of TI materials.

Low energy ion scattering (LEIS) is a simple, but powerful technique for surface structure analysis [21]. LEIS employs ions with incident energies between 0.5 and 10 keV, which have scattering kinematics that can be described with simple classical mechanics such that the energy of a singly scattered projectile is representative of the target atom mass. Due to shadowing and blocking, ions that travel deeper than a few atomic layers are not able to escape from the surface after only a single collision [21]. This makes LEIS especially surface sensitive and an ideal tool to determine the surface termination and atomic structure.

This paper compares surfaces prepared by *in situ* cleaving, *ex situ* cleaving and IBA. LEIS spectra are collected in a particular orientation to easily ascertain the composition of the outermost atomic layer, i.e., the termination. In addition, impact collision ion scattering spectrometry (ICISS) is used to provide information about the atomic structure of the top few layers. It is found that *in situ* cleaved and IBA-prepared surfaces are both Se-terminated, while *ex situ* cleaved samples can be either Bi-rich or Se-terminated.

**II. Experimental Procedure**

Single crystals with several stoichiometries, $Bi_2Se_{2.8}$, $Bi_2Se_{3.0}$, $Bi_2Se_{3.12}$ and $Bi_2Se_{3.2}$, were produced by a slow-cooling method [17,22]. The process involves melting mixtures of Bi shot (99.999%, Alfa Aesar) and Se shot (99.999+%, Alfa Aesar) in an evacuated quartz ampule (base pressure $\approx 2 \times 10^{-6}$ Torr) with an inner diameter of 17 mm. The mixture was heated to, and kept at, 750°C for one day, slowly cooled to 500°C for 68 hours, and then annealed at 500°C for three days before being cooled to room temperature. The sample ingot was then cleaved with a razor blade to obtain flat samples up to 10 mm in diameter. These samples naturally cleave along the (001) plane.

It was found that care must be taken in growing the single crystals to keep the contamination level low. In particular, the use of Bi shot, as opposed to powder, grows significantly better single crystals presumably because the high surface area of the powder contains more native oxide. In addition, the reactants need to be flushed with a high purity inert gas before being sealed in the ampule or the quality of the resulting single crystal will decrease.

Low energy electron diffraction (LEED) and LEIS measurements were conducted in an ultra-high vacuum (UHV) main chamber that has a base pressure of $2\times10^{-10}$ Torr. The main chamber is attached to a load-lock chamber and transfer system (Thermionics) that enables quick introduction of samples without the need for baking the main chamber. The foot of the manipulator contains an electron beam heater with a 0.008" thick tungsten filament located behind the sample holder. The main chamber includes a thermionic emission alkali ion gun (Kimball Physics), two detectors for performing LEIS, LEED optics (Princeton Research Instruments) and a sputter ion gun for sample cleaning.

For these experiments, the e-beam filament is used as a radiative heater without any applied bias voltage because the temperatures needed for annealing $Bi_2Se_3$ are rather low. As there is no thermocouple mounted to the transferable sample holder, the sample temperature is gauged via the filament current. The actual sample temperature was calibrated to the filament current by temporarily attaching a thermocouple to the center of an empty Ta sample holder. As this was done without a sample in place and the temperature can vary across the holder, the reported temperatures may be a bit larger than the actual sample temperature and the accuracy is limited to about 40°C.

LEIS was performed using an incident 3 keV $Na^+$ ion beam and the scattered projectiles were collected with either a time-of-flight (TOF) detector or an electrostatic analyzer (ESA). The ion beam is less than 1 mm in diameter. The ion gun and detectors are mounted in the same horizontal plane. The ion gun is mounted on a turntable that can rotate about the vertical axis of the chamber. The sample is mounted vertically on the manipulator which has two degrees of rotational freedom that allow the sample to rotate along the polar angle with respect to the ion gun and detectors, as well as azimuthally about its normal. As a result, the scattering angle, the incident polar angle and the outgoing azimuthal angle can all be changed independently. Note that the polar rotation utilizes a computer controlled stepper motor, so that measurements of the polar angular distribution of the ion yield are fully automated.

TOF uses a pulsed beam and measures the flight time of scattered ionic and neutral projectiles, as previously described [23]. The $Na^+$ ion beam is pulsed at 80 kHz by using deflection plates to pass the beam across a 1 mm aperture in the front of the ion gun. For the TOF measurements reported here, the ion beam is incident along the sample normal. The scattered particles are detected by a set of three micro-channel plates (MCP) mounted at the end of a 0.57 m long flight tube. There are two 3 mm diameter apertures in the flight tube, so that the acceptance angle for TOF spectra is less than 1°.

Energy spectra and angular distributions of the scattered ions are also collected with a 160° Comstock AC-901 hemispherical ESA that has a radius of 47.6 mm and a 2 mm diameter aperture, which makes the acceptance angle approximately 2°. The maximum scattering angle that can be obtained is 161º, which is the angle used with the ESA in this paper.

Although ion scattering is inherently a destructive technique, spectra can be collected before a significant fraction of the surface is damaged. The beam current is about 2 nA when using the ESA, but is reduced to about 10 pA when pulsing the beam for collecting TOF spectra. Thus, the collection of TOF spectra induces less beam damage than the collection of spectra with the ESA. For either detector, however, the incident ion fluence is kept below 1% of a monolayer so that the spectra are representative of the undamaged surfaces.

X-ray photoelectron spectroscopy (XPS) spectra were collected in a separate Kratos AXIS ULTRA XPS system equipped with an Al Kα monochromatic X-ray source and a 165 mm mean radius hemispherical ESA. The X-ray beam was incident at an angle of 60° relative to surface normal and has a spot size of about 1 mm in diameter. The main XPS chamber has a base pressure of $1\times10^{-9}$ Torr. It is also attached to a load-lock chamber that enables *in situ* cleaving and rapid sample introduction.

Sample surfaces are prepared by cleaving under vacuum (*in situ*), IBA or cleaving in air just prior to insertion (*ex situ*). After cleaving, the surfaces are visually flat and shiny. For *in situ* cleaving, the samples are attached to a stainless steel sample holder with Ag paste (Epoxy Technology). A vacuum-compatible epoxy (Accu-Glass Products, Inc.) is then used to mount an Al bar onto the sample. The pastes are cured by heating in a tube furnace in air at 100°C for 30 min. The samples are cleaved by knocking off the Al bar inside the UHV chamber. For *ex situ* cleaving, the sample is mounted onto the sample holder either with Ag paste or by Ta strips spot-welded onto the sample holder. Samples are then cleaved in air either with carbon tape or by knocking off the Al bar before being inserted into the load lock chamber. For IBA, the samples are always mounted onto a Ta sample holder with spot-welded Ta strips. They are cleaved *ex situ* before being introduced into the chamber and then prepared in UHV by repeated cycles of sputtering with 0.5 keV $Ar^+$ ions and annealing at 510°C for 30 min.

Simulations of ICISS data were performed using Kalypso, a windows-based software package that uses molecular dynamics (MD) to model atomic collisions in solids [24]. In these simulations, only the projectile-target atom repulsive interactions (screened Coulombic potential) are taken into account while the interactions between target atoms are ignored (the recoil interaction approximation). The Thomas-Fermi-Molière repulsive potential using the Firsov screening length is employed with a correction factor of 0.8 and a potential cut-off distance of 2.9 Å. The target model has four atomic layers ordered as Se-Bi-Se-Bi, with periodic boundary conditions used for the lateral planes. The specific atomic arrangement in the target includes surface relaxation by employing the average of the two sets of structural parameters obtained via LEED and SXRD in Ref. [10].

## III. Results

Experiments were performed using bismuth selenide samples synthesized with four different stoichiometries. The intent is to understand how the surface preparation method and stoichiometry affect the termination and atomic structure.

## A. LEED

Figure 1 shows LEED patterns collected from $Bi_2Se_{3.12}$ surfaces prepared by (a) *in situ* cleaving, (b) IBA, and (c) *ex situ* cleaving. All of the patterns were collected with the samples in the same position and using the same electron energy. Figures 1(a) and (b) each display bright and sharp 1x1 LEED patterns, indicating that both *in situ* cleaving and IBA lead to well-ordered crystalline surfaces. The pattern collected following *ex situ* cleaving was very dim, however, indicating a disordered surface.

In addition to providing the symmetry of the surface unit cell, LEED can also be used to determine the azimuthal orientation of the sample, although the pattern by itself is sometimes insufficient. For example, with $Bi_2Se_3$ the [120] orientation cannot be distinguished from the [210] orientation, as the symmetry of the LEED pattern is 6-fold while the symmetry of the crystal surface is 3-fold. Thus, LEIS spectra are used to identify the specific orientation, as explained below.

## B. XPS

Figure 2 shows Se 3d, Bi 4f and O 1s XPS spectra collected following both *in situ* and *ex situ* cleaving. Spectra for *in situ* cleaving were collected from a $Bi_2Se_{3.12}$ sample that was cleaved under vacuum in the load lock and then transferred into the XPS main chamber. The absence of an O 1s signal and the presence of only a single core level component in the Bi 4f and Se 3d XPS spectra demonstrate that the samples following *in situ* cleaving were without contamination.

*Ex situ* cleaving was performed using carbon tape for one $Bi_2Se_{3.12}$ sample and two $Bi_2Se_{3.0}$ samples. For each measurement, the three samples were cleaved, exposed to air for 5 min, 1 hour or 1 day, and then inserted into the chamber together. All three samples gave the same results, so only the data for $Bi_2Se_{3.12}$ is shown here. The data collected following *ex situ* cleaving display oxygen contamination. The Se and Bi core levels show shifted components consistent with Se-O and Bi-O bonds, similar to what was reported in ref. [15], and there is a clearly visible O 1s core level. Note that the C 1s core level is not visible even after air exposure for 1 day, as it overlaps with the Se $L_2M_{23}M_{45}$ Auger line making it difficult to discern. The signals for the Se-O and Bi-O components and the O 1s level all increase with air exposure, indicating that the surfaces oxidize gradually over time. The Bi-Bi metallic bonds that were sometimes observed following *ex situ* cleaving in refs. [9,15] were not detected. This is not unexpected, however, considering that the probability for an *ex situ* cleaved sample to be Bi-rich is small and there were only three trials in which XPS spectra were collected here.

**C. Time-of-flight Low Energy Ion Scattering**

LEIS involves the bombardment of a surface with keV ions and the collection and energy analysis of the projectiles that scatter back from the sample. Ions in this energy range can be modeled with the binary collision approximation (BCA), which assumes that the projectiles make a series of isolated binary collisions with unbound surface atoms located at the lattice sites. This is a reasonable assumption as the scattering cross sections are generally smaller than the spacings between atoms and the bonding energy of the surface atoms is much smaller than the projectile's kinetic energy. If a projectile backscatters after colliding with only a single surface atom, then the energy of the scattered projectile is primarily a function of the projectile/target mass ratio and the scattering angle [21]. Thus, LEIS spectra display a single scattering peak (SSP) for each element on the surface that is directly visible to both the incident ion beam and the detector.

The intensity of the single scattering yield as a function of angle also depends on the particular atomic arrangement in the outermost two or three layers. The yield for a particular SSP depends primarily on the number of such atoms that are visible, but is also affected by focusing that occurs when the projectile experiences a grazing collision with a surface atom, which involves very little energy loss, and subsequently undergoes a hard collision with another atom that leads to backscattering at the SSP energy. Thus, the yield can change greatly as the specific orientation between the incident ions, the single crystal sample and the detector is adjusted. It is the thus angular dependence of the scattering yield that is measured when using LEIS to determine a surface structure, and there are a number of ways in which this can be accomplished.

For example, the orientation can be set in such a way as to make some near-surface atoms visible to the incoming ion beam, while keeping others hidden. An example of this is illustrated schematically in Fig. 3(a), which shows a side view of the nominal bulk terminated $Bi_2Se_3$ surface along the [210] azimuthal direction. The figure depicts an incident ion beam directed along the surface normal and the "shadow cones" formed by interaction with atoms in the first three atomic layers. Such shadow cones are constructed by mapping out the possible trajectories of the incoming ions [25]. For $Bi_2Se_3$, this means that the normally incident beam can only directly impact atoms in the first three atomic layers, as the fourth, fifth and sixth layer atoms are shadowed by the first, second and third layer atoms, respectively. Such a geometry in which the incident beam is aimed along a low index crystal direction is referred to as "single alignment".

A "double alignment" orientation occurs when the detector is also aligned along a low-index crystal direction. As an example, placing the detector along the bond angle between second layer Bi and first layer Se, which is at ~33° from the surface plane in the bulk-terminated material, leads to a geometry in which the second layer atoms block projectiles scattered from the third layer from directly reaching the detector, as illustrated in Fig. 3(a). Similarly, projectiles that impact the second layer are blocked from reaching the detector by the first layer atoms. There are "blocking cones" illustrated in the figure, which are similar to shadow cones but are formed by the possible trajectories of ions that are initially scattered from a deeper layer atom. Thus, in this double alignment orientation, only the outermost atomic layer contributes to the SSP. Note that since the diameter of shadow and blocking cones are on the order of Å's for low energy ions, a scattering angle that is a few degrees off from the actual bond angle is sufficient to maintain the double alignment orientation. Ion scattering spectra collected in such a double alignment orientation are thus an ideal tool for determining the composition of the outermost atomic layer, i.e., the surface termination.

Figure 4 shows TOF spectra collected in double and single alignment orientations from samples prepared by *in situ* cleaving and IBA. A single alignment orientation occurs by maintaining the polar angle of 33° but rotating the sample azimuthally about its normal so that the surface atoms no longer block projectiles that initially hit the 2nd or 3rd layer atoms. The spectra shown in Figs. 4(b) and (d) were collected by rotating the samples azimuthally 60° from the double alignment orientation so that the exit direction is now along the [120] projection. The difference between these two orientations is used here to distinguish the [210] and [120] azimuthal directions, as the LEED pattern alone cannot do this. LEIS spectra are collected for each orientation, and the one that shows only a single SSP is identified as the [210] direction.

The feature at 6.3 μs in Fig. 4 is the Se SSP and the feature at 4.8 μs is the Bi SSP, which is consistent with the expectation that the particles scattered from heavier target atoms will have a larger kinetic energy and thus reach the MCP detector more quickly. The large backgrounds that extend from about 9 μs to flight times just beyond the Bi SSP's are due to multiple scattering trajectories. This type of background is present in all of the TOF spectra. Note that the actual background increases at longer flight times due to the cascade of multiply scattered particles, but the transmission function of the detector goes down quickly with lower impact energies because of the decreasing MCP efficiency so that the background in the experimental data approaches zero at the longest flight times [26].

The TOF spectra collected in the double alignment orientation provide the terminations of the surfaces, as only the outermost atomic layer contributes to the SSP. Figure 4(a) shows a spectrum collected from an *in situ* cleaved surface. This spectrum shows a clear Se SSP riding on the background, and no Bi SSP, which indicates a Se-termination. Figure 4(c) was collected from an IBA prepared surface, and also indicates a Se-termination. The single alignment spectra in Figs. 4(b) and 4(d) show both Se and Bi SSP's, indicating the presence of Bi in the second or perhaps third layer.

*Ex situ* cleaved samples have a less reproducible behavior, sometimes showing Se-termination and other times being Bi-rich. Spectra collected after *ex situ* cleaving that display each of these possibilities are shown in Fig. 5. The *ex situ* cleaved samples were exposed to air for less than 5 min before being placed into the load lock chamber. The spectra shown in Figs. 5(a) and (b) are similar to those collected from *in situ* cleaved and IBA surfaces, suggesting that this sample is Se-terminated with Bi in the second or third layer. Figures 5(c) and (d) show a Bi SSP in both double and single alignment, indicating that there is Bi in the outermost surface layer. The noise level in these spectra is rather large, however, making it difficult to discern a Se SSP. Thus, it's not clear from the LEIS spectra whether the surface is terminated with an ordered Bi layer or if it is covered by islands of Bi metal that obscure much of the Se signal. Because the LEED pattern associated with *ex situ* cleaved samples is weak, the latter possibility is more likely, considering that Bi is a strong electron scatterer that would be less affected by adsorbates than an ordered Se-termination would be. Thus, this surface is designated as Bi-rich rather than Bi-terminated.

The data in Figs. 5(a) through (d) also show evidence for adsorbed contaminants on the *ex situ* cleaved surfaces. First, the spectra all have a shoulder-like feature between 2.2 μs and 4.5 μs, which is highlighted in Fig. 5(a). This shoulder results from direct recoiling of light adsorbed atoms, such as hydrogen, carbon or oxygen. Direct recoiling occurs when an atom is removed from the surface in a collision that converts most of the projectile's kinetic energy to the kinetic energy of the recoiled atom [21]. Direct recoiling is most pronounced for large projectile/target mass ratios, and such fast recoiled atomic particles have sufficient kinetic energy to induce a signal in the MCP. Note that direct recoiling is different than sputtering. In sputtering, the emitted atoms result from a collision cascade and have less than 10 eV of kinetic energy, so they would not be counted by the MCP detector. Additional evidence for the presence of adsorbates is given by the Se SSPs, which are smaller relative to the multiple scattering background than for an *in situ* or IBA-prepared surface, which is likely caused by surface contamination partially covering the outermost Se atoms. Surface contamination would also explain why the LEED pattern for a sample cleaved *ex situ* is always very dim. It is also possible that the contamination could be a factor leading to the non-reproducibility of the surface termination following *ex situ* cleaving, as discussed below.

The *ex situ* cleaved Se-terminated surface is further studied by subjecting it to a mild annealing in vacuum. Figure 6 shows TOF spectra collected following *in situ* cleaving, and then after annealing at 130°C and 290°C for 30 minutes each. The data show that annealing largely reduces the feature associated with direct recoiling and increases the size of the Se SSP. The LEED pattern after annealing (not shown) becomes as sharp and bright as that following *in situ* cleaving. These LEIS and LEED results thus suggest that the *ex situ* cleaved surface is similar to the Se-terminated surface formed by *in situ* cleaving, but is covered by contaminants that can be removed by a light anneal.

Different stoichiometries of bismuth selenide are used to investigate if excess Bi or Se affects the surface termination. Although all of the TOF spectra shown in Figs. 4 and 5 were collected from Bi$_2$Se$_{3.12}$, they are indistinguishable from spectra collected from stoichiometric Bi$_2$Se$_3$. Table 1 shows the number of times that each type of surface was produced when using different stoichiometries and preparation methods. As the table shows, *in situ* cleaved and IBA-prepared surfaces were always Se-terminated for all of the stoichiometries tested. Out of 17 trials of *ex situ* cleaving, 13 were Se-terminated and 4 were Bi-rich. The probability for obtaining a Bi-rich surface, averaged over all of the stoichiometries, is thus around 22%. Bi$_2$Se$_{3.12}$ and Bi$_2$Se$_{3.0}$ have each produced Bi-rich and Se-terminated surfaces following *ex situ* cleaving. Thus, the termination does not appear be to sensitive to the bulk stoichiometry, as far as this limited numbers of trials can demonstrate.

**C. Impact Collision Ion Scattering Spectrometry**

ICISS is a variation of LEIS that provides information about the atomic structure of the outermost few layers of a single crystal solid by monitoring the yield of projectiles that are singly scattered at a large angle as the sample is rotated [27,28]. The basic idea is that the projection of the shadow cones formed by atoms in the outermost layer changes with respect to other surface and near-surface atoms. The flux of projectiles is zero within a shadow cone, but is peaked at the edges of the cone. Thus, the yield of projectiles singly scattered from the surface and near-surface atoms changes in response to this change in flux as the sample rotates. The advantage to the large scattering angle employed for ICISS is that the effects of the shadow and blocking cones can be considered separately and in terms of two-atom pairs within the atomic structure. With a smaller scattering angle, particular trajectories are more likely to be influenced by multiple cones, making the analysis more complex. The ESA is used for ICISS, rather than TOF, as it can be set to the energy of a particular SSP and that signal then monitored as the incident angle is adjusted. In doing so, it is implicitly assumed that neutralization does not affect the shape of the angular yields.

Figure 7 shows an energy spectrum collected from $Bi_2Se_{3,12}$ with the ESA. The peak at 0.8 keV is the Se SSP and the peak at 1.8 keV is the Bi SSP. The background is considerably different than in the TOF spectra because the ESA measures only the ion yield and the transmission function does not change with scattered energy as with the TOF detector. This spectrum is shown because it is used to identify the Se and Bi SSP peak positions that are used for ICISS polar angle scans.

Figure 8 shows ICISS polar scans collected along the [120] azimuth from an IBA-prepared and an *in situ* cleaved surface. The features in these scans can be qualitatively understood by considering how the shadow and blocking cones associated with one atom interact with a second atom in the crystal structure as the sample rotates. The schematic diagram in Fig. 9 shows four such two-atom pairs and helps to illustrate the angles at which three shadow cones and one blocking cone act to sharply increase or decrease the SSP yield, assuming a Se-terminated structure. There will be no backscattered yield when the incident beam is just at or barely above the surface plane, as each surface atom is inside the shadow cone of its neighbor. As the incident angle increases towards the surface normal, the edge of the shadow cone due to a first layer atom will pass through a neighboring first layer atom, as illustrated by the 1s trajectory in Fig. 9. When this happens, the Se SSP intensity will display a maximum, due to the increased flux at shadow cone edge, and then decrease to the steady value consistent with unimpeded scattering from the surface atoms. A feature that results from the shadow cone of a first layer atom passing through a neighboring first layer atom is called a surface flux peak (SFP). The peak at 12° in the Se SSP angular scan in Fig. 8(b) is the SFP for the Se-terminated structure. Additional features arise in the ICISS scans as the shadow cones around the first layer atoms interact with atoms in deeper layers. For example, the peak at 55º for the Bi SSP in Fig. 8(a) occurs when the shadow cone of the outermost Se atom interacts with a second layer Bi atom, as illustrated by trajectory 2s in Fig. 9. In a similar way, the broad feature at approximately 60º in Fig. 8(b) has contributions from the enhancement that occurs when the shadow cone of a second layer Bi atom interacts with third layer Se, as illustrated by trajectory 3s. When the scattering angle is less than 180°, blocking can also contribute features to an ICISS scan. The decrease of at 86° in Fig. 8(b) is due to the blocking by first layer Se atoms of projectiles scattered from the third layer, as illustrated by trajectory 4b in Fig. 9. A detailed analysis of ICISS data from Se-terminated surfaces at several azimuths compared with simulations will be published separately [29].

The peak positions in the ICISS angular scans are strongly related to the crystal structure, so that the good agreement between the polar scans from the IBA-prepared and *in situ* cleaved surfaces confirms that the near-surface atomic structures are the same. Furthermore, this analysis of the ICISS data indicates that the top three layers are ordered as Se-Bi-Se, which is consistent with an intact QL at the surface.

To get additional confirmation that IBA and *in situ* cleaved surfaces are both Se-terminated,

ICISS simulations were performed using Se-terminated and single layer Bi-terminated models, with the results shown in Fig. 10. The simulations are confined within the ($\bar{1}20$) plane, as illustrated in Fig. 9, which greatly reduces the computation time. This simplification works well because the distance between ($\bar{1}20$) planes, 2.07 Å, is larger than the sizes of blocking and shadow cones so that single scattering trajectories along the [120] azimuth in different ($\bar{1}20$) planes are completely independent. Both models have four atomic layers with four atoms in each layer, and a periodic boundary condition is applied along the [120] direction. The Bi-terminated model is assumed to consist of Bi-Se-Bi-Se, with the structural parameters all the same as the Se-terminated model. The experimental data shown as a solid line in Fig. 10 are the same as shown in Fig. 8(a). The experimental and simulated data both display peaks at around 55°, indicating $2^{nd}$ or $3^{rd}$ layer Bi. The simulation for a Bi-termination has a SFP at 9°, however, which is missing in the experimental data. Thus, the simulations strongly rule out the possibility that the IBA-prepared surface is Bi-terminated. This agrees with the conclusions drawn from the TOF spectra in Fig. 4 and the ICISS data in Fig. 8.

**IV. Discussion**

The fact that *in situ* cleaving and IBA both produce a well-ordered Se-terminated surface suggests that this is the stable configuration, in accordance with the expectations for such a layered material. When $Bi_2Se_3$ is cleaved *ex situ*, a contaminant-covered Se-terminated surface is sometimes produced, while at other times a contaminant-covered Bi-rich surface results. Thus, it is also reasonable to conclude that cleaving always produces a Se-terminated surface, but changes can occur after cleaving. Note that this was the conclusion of ref. [12], although that paper reported that the resulting surface was Bi-terminated and not simply Bi-rich. Post-cleavage changes to the surface composition and structure for some *ex situ* cleaved samples could result from a surface chemical reaction with atmospheric contaminants, and such contaminants are likely to preferentially adsorb at surface defects.

Defects in these materials are typically Se vacancies, as Se dimers and tetramers are stable gas phase species that can desorb from the surface. Se vacancies are often present in the as-grown materials and are generally thought to be the reason for the natural n-type doping of $Bi_2Se_3$ [18,20,30,31].

The stress applied to the sample during cleaving can also produce defects. The propensity of defects to form during cleaving depends on many factors, such as the crystal quality and the method used to cleave. After a single crystal is grown, a razor blade is commonly used to break the crystal along the natural (001) cleavage plane, and then cleaving with tape or paste is used to produce a flat surface. Cleaving induces strain in the surface, which might then alter the surface atomic structure and the TSS. For example, mechanical strain in TI's was predicted by DFT to affect the electronic properties by changing the Γ point band gap [32], which was later verified experimentally using the strain existing at the surface grain boundaries of a $Bi_2Se_3$ film [33,34]. Another indirect example of the creation of defects comes from studies of cleaved Si(111) surfaces, which showed that stress-induced microstructures with two types of terraces, triangular terraces and parallel-steps terraces, can coexist on the same surface [35]. Because the process and the force applied on the samples during cleaving are not well controlled, it is not surprising that the density of defects is not reproducible from sample to sample.

Different cleaving methods can also lead to different defect concentrations. Prior to usage, each $Bi_2Se_3$ piece is usually cleaved with tape several times in an attempt to produce a flat surface and reduce the number of defects. Reference [36] claims, however, that cleaving with scotch tape can still induce many defects, while using a mechanical cleaver can produce a nearly atomically flat surface. Triangular defects with half height QL step edges induced by scotch tape cleaving were also shown in Ref. [37].

The sample size is also important. If the sample is larger, the center will be relatively less affected by the stress during cleaving and thus be more defect free. The center of larger samples is also better protected from contamination during handling. These considerations may explain why the samples used in Refs. [14,36] were very flat and inert, as the Bi metal was purified before usage, the samples were more than 1 cm in diameter and they were prepared with a mechanical cleaver. In contrast, most of the experiments with $Bi_2Se_3$ reported in the literature used samples smaller than 5 mm and were cleaved with tape [12,38].

Chemical reactions of atmospheric contaminants with surface defects are a likely cause of the surface structural changes that occur with some *ex situ* cleaved samples. For example, in synthesizing the materials for the present study, it was found that $Bi_2Se_3$ is very sensitive to oxygen contamination. When samples are cleaved in UHV or in a glove box filled with an inert gas, the newly cleaved surfaces have no contact with atmospheric gases. If a chemical reaction was responsible for the termination change, this would explain why these remained

Se-terminated. The detection of oxygen on *ex situ* cleaved surfaces, as well as data widely found in the literature, indicates the possibility of a surface reaction with air [13,15,39]. In addition, scanning photoelectron microscopy shows that step edges oxidize much faster than the basal planes for $Bi_2Te_2Se$ [39], leading to the conclusion that step edge densities will have a profound effect on the rate of oxidation.

Based on the above analysis, it is proposed that the Bi-rich surface is formed by chemical reaction of adsorbed contaminants with a Se-terminated surface that has a high defect density. Atmospheric gases, such as water or oxygen, can react with the defects to somehow decrease the surface concentration of Se. For example, oxygen might break a Bi-Se bond and release volatile Se to leave a Bi-rich surface.

If there are too many defects, the Bi-rich surface might lead to an actual Bi-termination as found in our previous LEIS study [12]. The as-grown samples used in Ref. [12] were small, on the order of 3 mm in diameter. Smaller samples will have more surface defects and smaller surface basal planes, as mentioned above. This explanation is also supported by the fact that the TOF spectra collected after *in situ* cleaving of the small samples used in Ref. [12] usually had obvious recoiling shoulders. Although the TOF spectra that were published in Ref. [12] do not have a noticeable recoiling feature, this doesn't mean there was absolutely no contamination. Instead, this might indicate that only a small concentration of contaminants is actually needed to alter the surface structure, which is consistent with the rapid oxidation widely observed in the literature [13]. Thus, even small amounts of contamination from background gases in the UHV chamber, or from the sample holder itself, might react with a high density of surface defects to alter the termination.

For the present experiments, large samples that are around 10 mm in diameter were grown. In addition, the Bi and Se shot were flashed twice with high purity argon before being sealed in the high vacuum ampule to minimize the intrinsic contamination and density of defects. These improvements helped to reduce the contamination level and the number of defects so that the samples used here are much more likely to remain Se-terminated after cleaving.

**V. Conclusions**

*In situ* cleaved and IBA-prepared surfaces are both well-ordered and terminated as Se-Bi-Se, which is the expected structure assuming that the surfaces cleave along the van der Waals gap to reveal an intact QL. Surfaces cleaved *ex situ*, however, are covered with a submonolayer of contaminants and can be either Se-terminated or Bi-rich. This contamination may be involved in a chemical reaction that is ultimately responsible for the non-reproducibility of the surface termination. It is proposed that defects on the surface would increase the adsorption of contaminants and thus the propensity for a termination change. Samples that were cleaved *ex situ* can be returned to a Se-termination by an IBA process, whether they had been Se-terminated or Bi-rich. Thus, IBA would be the preferred method for producing a high quality surface for UHV studies. Investigations of the surface defects produced by cleaving and studies of the chemical reactions of atmospheric contaminants with the defects are needed to fully understand the chemistry of *ex situ* cleaved surfaces, how such reactions can sometimes lead to a Bi-rich surface and how to develop robust methods of sample preparation.

## VI. Acknowledgements

The authors would like to thank Dr. Zhiyong Wang and Prof. Ian Fisher for instructing us about the growth of $Bi_2Se_3$ single crystals, and Dr. M.A. Karolewski for help with Kalypso. This material is based on work supported by, or in part by, the U.S. Army Research Laboratory and the U.S. Army Research Office under Grant No. 63852-PH-H.

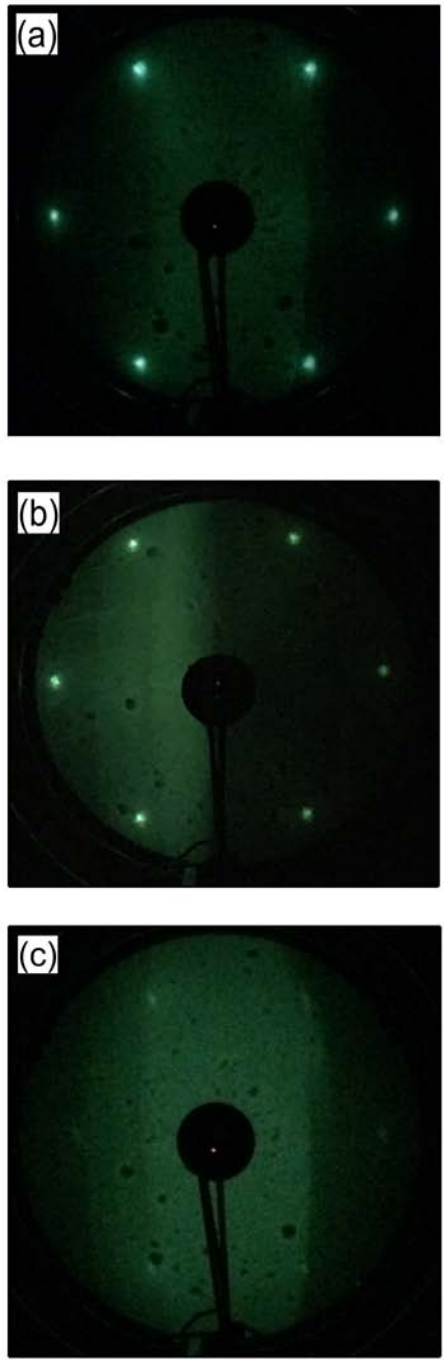

**Figure 1.** LEED patterns collected with an electron kinetic energy of 25.9 eV from $Bi_2Se_3$.[12] surfaces prepared by (a) *in situ* cleaving, (b) IBA, and (c) *ex situ* cleaving.

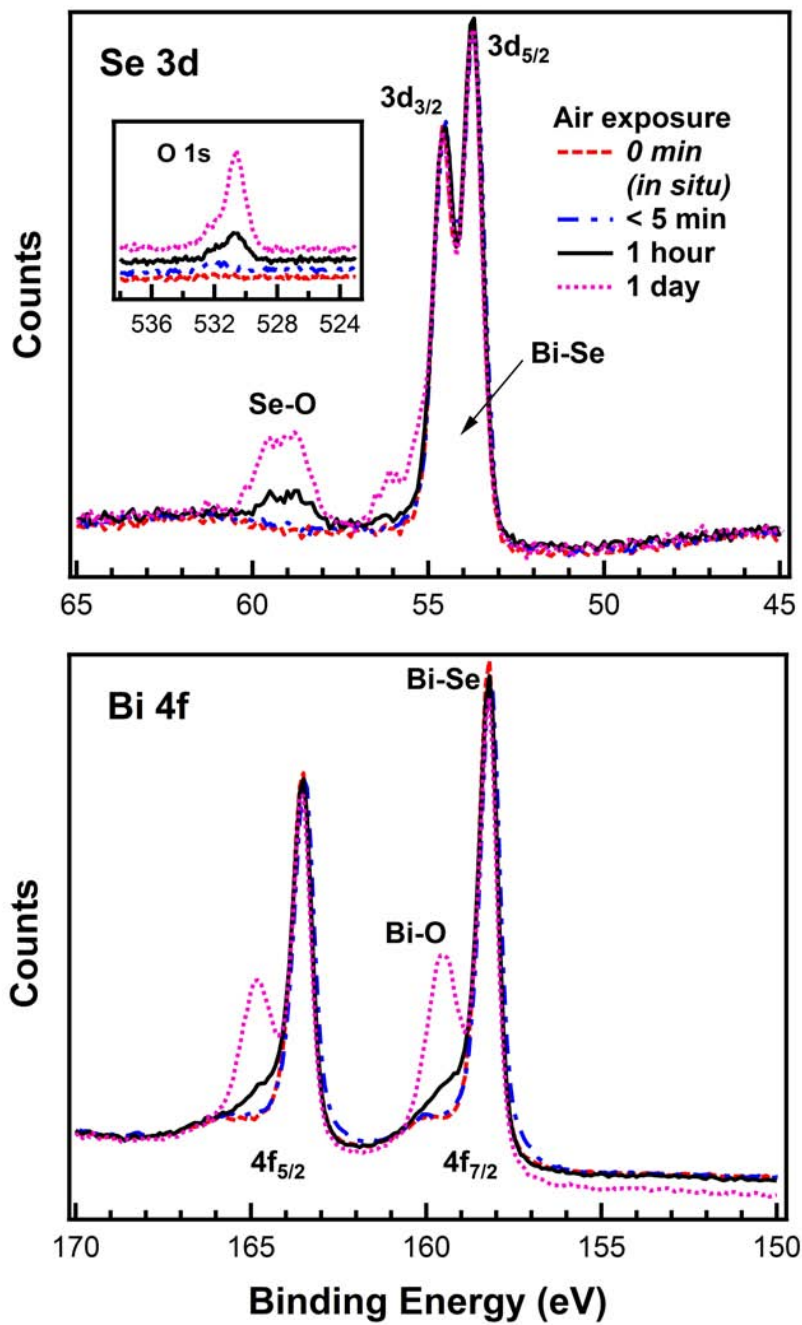

**Figure 2.** Se 3d and Bi 4f XPS core level spectra collected from an *in situ* cleaved Bi$_2$Se$_3$.$_{12}$ surface and an *ex situ* cleaved Bi$_2$Se$_3$.$_{12}$ surface exposed to air for less than 5 min, 1 hour and 1 day. The inset in the upper panel shows the O 1s level.

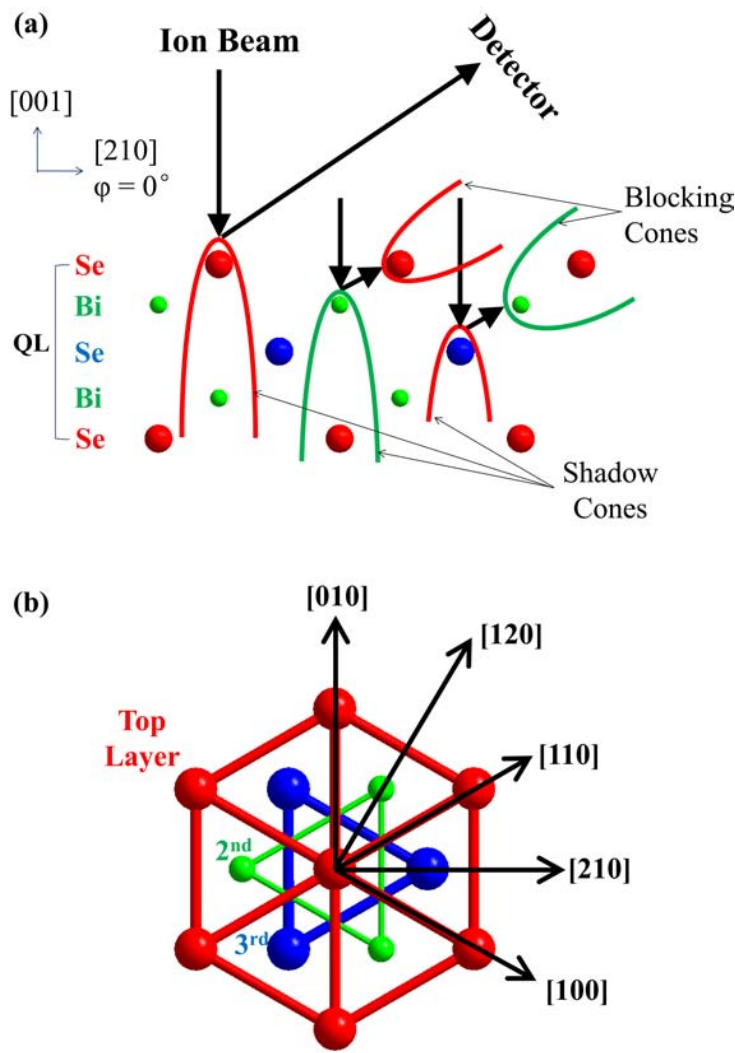

**Figure 3.** Schematic diagrams indicating the orientation used in collecting TOF spectra from $Bi_2Se_3$ surfaces with the incident ion beam normal to the sample surface. Diagram (a) illustrates a side view in the double alignment orientation with the detector is positioned at an angle of 33° from the surface plane. Diagram (b) illustrates a top view of the sample surface and various low index azimuthal directions, including the outgoing [210] azimuth used for the double alignment orientation. The models are shown using ionic radii with accurate atomic positions.

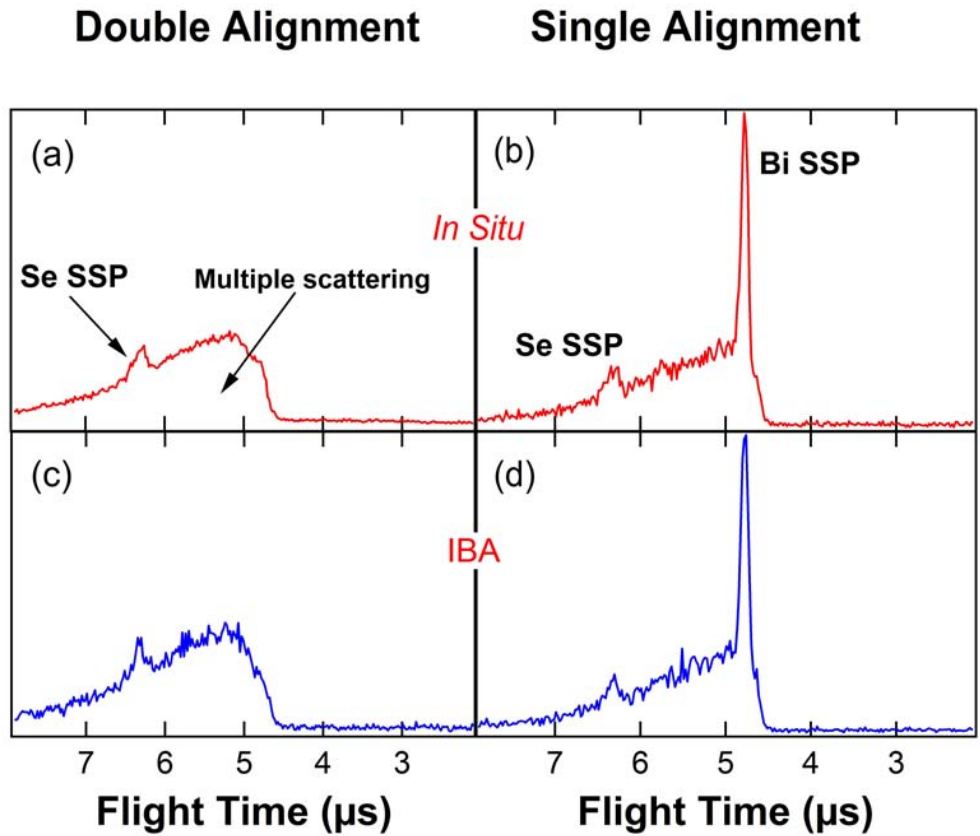

**Figure 4.** TOF spectra collected using normally incident 3.0 keV Na$^+$ ions scattered at 126° from Bi$_2$Se$_{3.12}$ surfaces prepared by *in situ* cleaving (a, b) and IBA (c, d). The spectra on the left were collected in double alignment orientation in which the outgoing projectiles are along the [210] azimuthal projection, while the spectra on the right were collected using the [120] azimuthal projection.

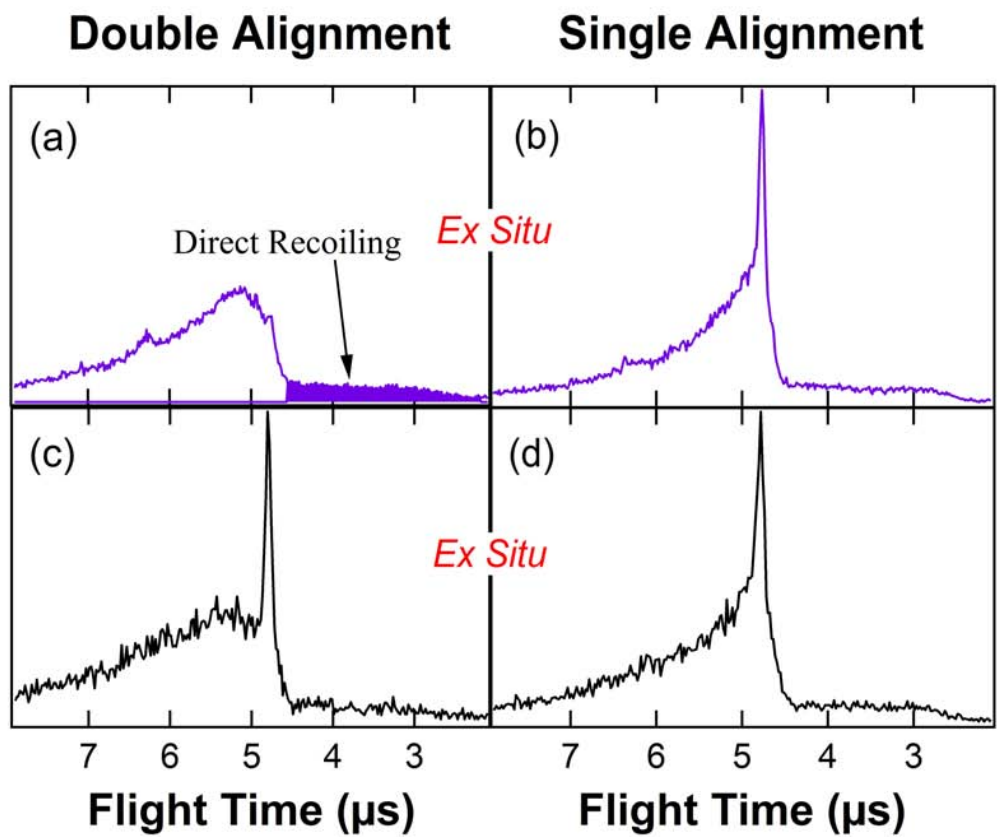

**Figure 5.** TOF spectra collected using normally incident 3.0 keV Na$^+$ ions scattered at 126° from Bi$_2$Se$_{3.12}$ surfaces prepared by *ex situ* cleaving. The spectra on the left were collected in double alignment orientation in which the outgoing projectiles are along the [210] azimuthal projection, while the spectra on the right were collected using the [120] azimuthal projection.

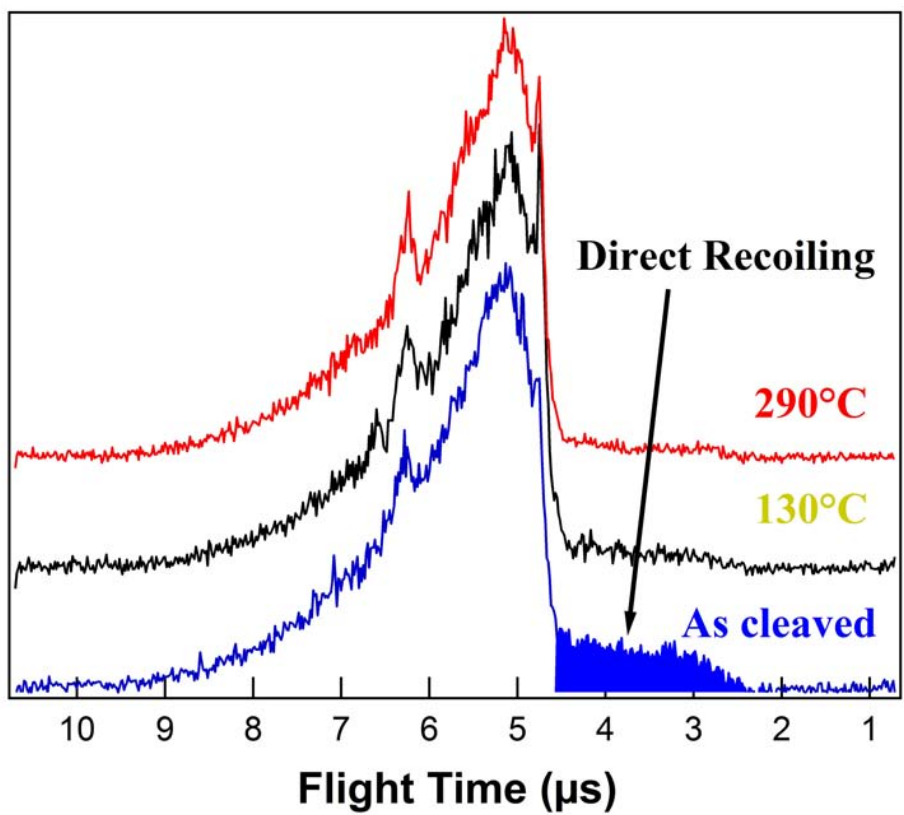

**Figure 6.** TOF spectra collected using normally incident 3.0 keV Na$^+$ ions scattered at 126° from Bi$_2$Se$_{3.12}$ surfaces following *ex situ* cleaving and annealing at 130°C and 290°C. The bottom spectrum is the same as that shown in Fig. 5(a).

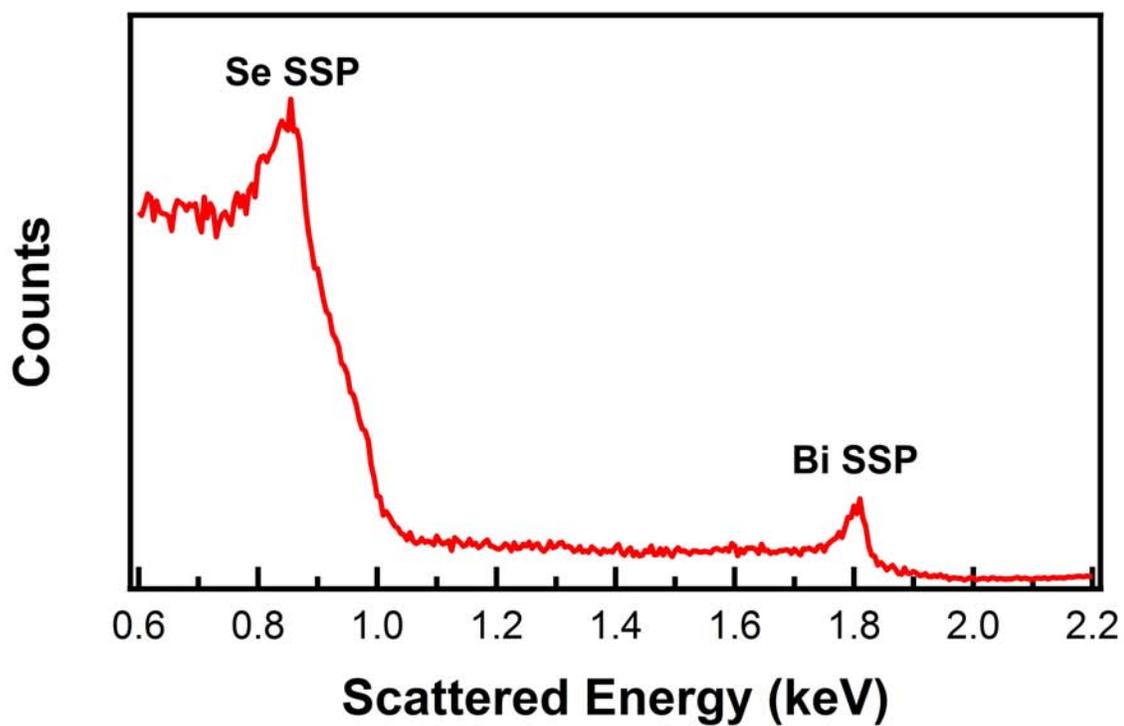

**Figure 7.** Energy spectrum collected with the ESA for 3.0 keV Na$^+$ scattered from an *in situ* cleaved Bi$_2$Se$_{3.12}$ surface with an incident polar angle of 86° from the surface plane along the [010] azimuth using a scattering angle of 161°.

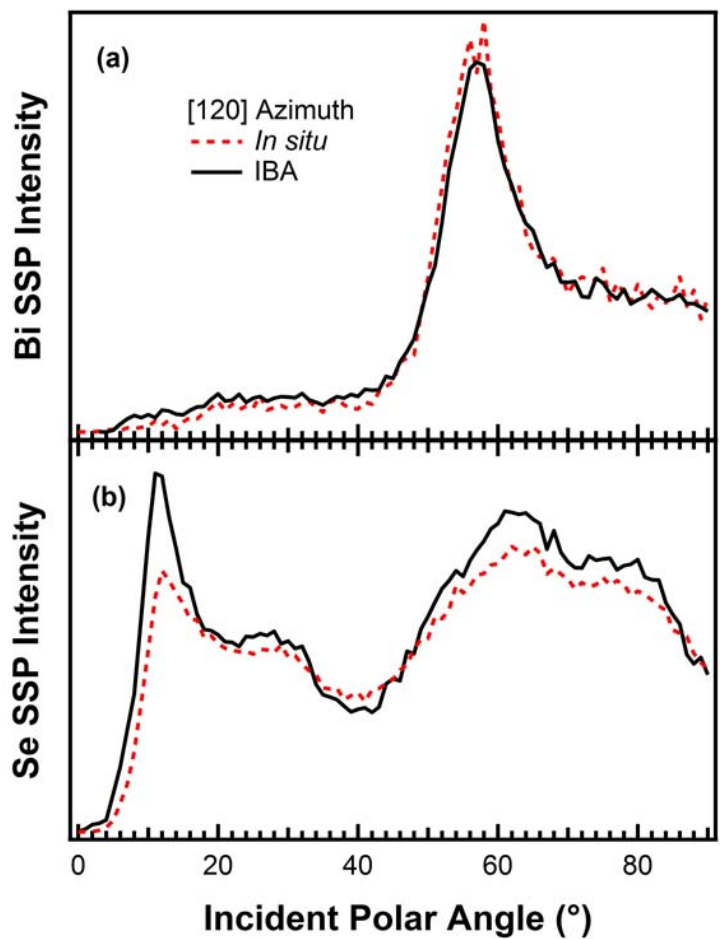

**Figure 8.** The polar angle dependence of the intensities of (a) the Bi SSP and (b) the Se SSP for $Bi_2Se_{3.12}$ surfaces prepared *in situ* (dashed line) and by IBA (solid line) collected along the [120] azimuth using a scattering angle of 161°. The intensities are adjusted to match at a 90° polar angle.

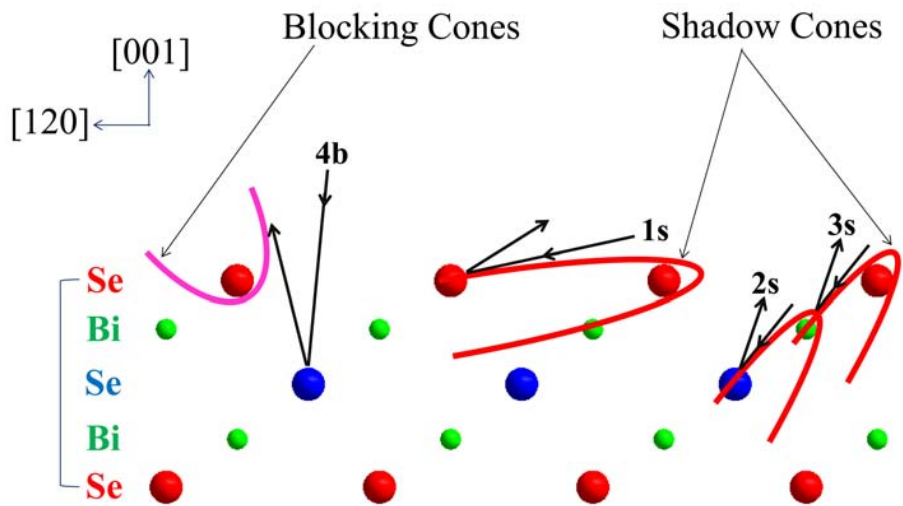

**Figure 9**. A side view of the ($\bar{1}$20) plane containing the [120] azimuth. The arrows show four primary trajectories that contribute to the experimental ICISS polar scans along the [120] azimuth for a scattering angle of 161°. The letter *s* refers to a shadowing alignment and the *b* to a blocking alignment.

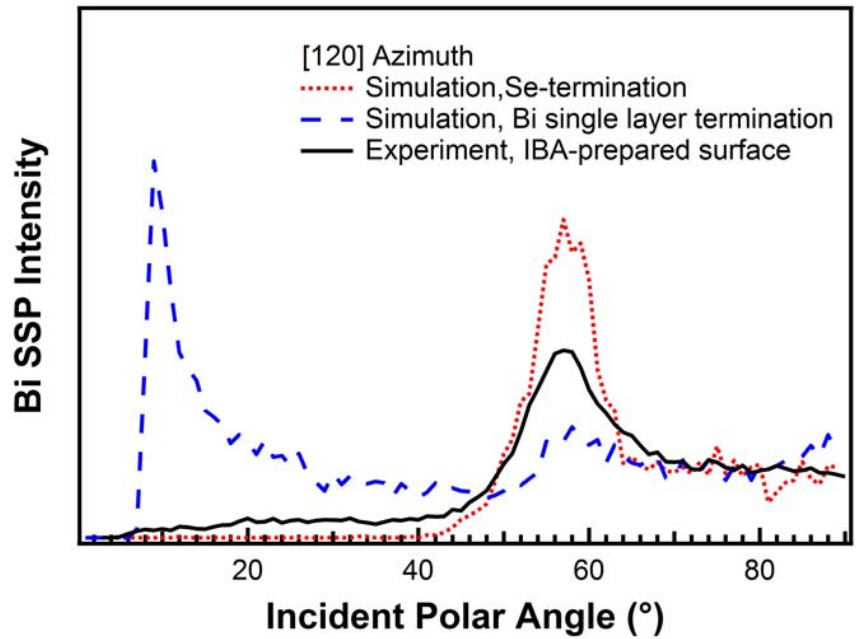

**Figure 10.** The polar angle scan of the Bi SSP intensity along the [120] azimuth using a scattering angle of 161°. The graph shows experimental data (solid line) and simulations using models with Se-termination (short dashed line) and Bi single layer termination (long dashed line). Spectra are adjusted to make the intensity at the polar angle of around 90° match.

**Table 1**. The number of times that each type of surface termination resulted when using $Bi_2Se_3$ with different stoichiometries and employing the various surface preparation methods.

|  |  | $Bi_2Se_{2.8}$ | $Bi_2Se_{3.0}$ | $Bi_2Se_{3.12}$ | $Bi_2Se_{3.2}$ |
|---|---|---|---|---|---|
| *In Situ* | Se-termination | 0 | 9 | 10 | 2 |
| *Ex Situ* | Bi-rich | 0 | 1 | 3 | 0 |
|  | Se-termination | 2 | 7 | 4 | 1 |
| **IBA** | Se-termination | 0 | 1 | 5 | 0 |